\documentclass[showpacs,twocolumn,preprintnumbers,superscriptaddress,amsmath,amssymb,nofootinbib]{revtex4}
\usepackage{graphicx,color}
\usepackage[section]{placeins}
\newcommand{\cd}{\makebox[0.08cm]{$\cdot$}}

\newcommand{\sla}{\not\!}

\begin{document}
\vspace{0.5cm}
\pagestyle{headings}

\title{Taylor-Lagrange renormalization scheme, \\ Pauli-Villars subtraction and Light-Front dynamics}

\author{P. Grang\'e}
\affiliation{Laboratoire de Physique Th\'eorique et Astroparticules,\\ Universit\'e Montpellier II, CNRS/IN2P3, Place E. Bataillon\\ F-34095 Montpellier Cedex 05, France}
\author{J.-F.~Mathiot}
\affiliation {Clermont Universit\'e, Laboratoire de Physique
Corpusculaire, \\ BP10448, F-63000 Clermont-Ferrand, France}
\author{B. Mutet}
\affiliation{Laboratoire de Physique Th\'eorique et Astroparticules,\\ Universit\'e Montpellier II, CNRS/IN2P3, Place E. Bataillon\\ F-34095 Montpellier Cedex 05, France}
\bibliographystyle{unsrt}
\author{E. Werner}
\affiliation{Institut f$\ddot u$r Theoretische Physik, Universit$\ddot a$t Regensburg,\\ Universit$\ddot a$tstrasse 31, \ D-93053 Regensburg,  Germany}

\begin{abstract}
We show how the recently proposed Taylor-Lagrange renormalization scheme can lead to extensions of singular distributions which are reminiscent of the Pauli-Villars subtraction. However, at variance with  the Pauli-Villars regularization scheme, no infinite mass limit is performed in this scheme. As an illustration of this mechanism, we consider the calculation of the self-energy in second order perturbation theory in the Yukawa model, within the covariant formulation of light-front dynamics. We show in particular how rotational invariance is preserved in this scheme.
\end{abstract}
\maketitle

%%%%%%%%%%%%%%%%%%%%%%%%%%%%%%%%%%%%%%%%%%%%%%%%
\section{Introduction}
The understanding of the structure of bound state systems in nuclear and particle physics requires the development of a relativistic nonperturbative framework. For obvious practical reasons, any calculation of this type relies on approximations, and one thus needs a systematic strategy in order to improve the approximations which are made, in complete analogy with perturbation theory.

 Light-front dynamics (LFD) is a very powerful tool  to calculate bound state properties. It is one of the three forms of dynamics proposed in 1949 by Dirac \cite{dirac}. 
In order to perform systematic calculations of physical observables on a large scale, one should however be able to solve three important problems.

The first one is the explicit violation of rotational invariance by the choice of a given light-front plane. The control of this violation is important in order to define unambiguously all physical observables. This can be done simply using the covariant formulation of light-front dynamics (CLFD) \cite{karmanov,cdkm}.

The second one is the possible appearance of uncanceled divergences when the Fock space is truncated, in any approximate nonperturbative calculation. One thus should make sure that no divergences are left uncanceled. This is enforced using the Fock sector dependent renormalization scheme \cite{kms_08}.

Finally, one should develop a regularization/renormalization scheme which preserves all symmetries, and which is well adapted to extended numerical calculations.
 
We have advocated in a previous study \cite{grange} the use of the Taylor-Lagrange renormalization scheme (TLRS) \cite{GW}. We have shown in 
particular that this scheme is very well adapted to any calculation in LFD. It is systematic, can treat singularities of any type on 
the same footing, and moreover does not require to perform any infinite scale limit.

Many other regularization methods are available in the literature. Let us mention here the most important ones.

{\it i) The cut-off method}. This is a simple, but to some extent brutal, way to regularize divergent amplitudes. It however violates gauge invariance and, in LFD, rotational invariance \cite{kms_04}. It should thus be avoided in any realistic calculation.

{\it ii) The Pauli-Villars (PV) subtraction method}. In LFD, the PV regularization scheme amounts to extend the Fock space to include PV fields with negative norm \cite{kms_08}. While this method is attractive for its simplicity and immediate use, it has some serious disadvantages in any systematic calculations. The number of PV fields may be large if singularities are of high order, as in effective field theories. This may imply a large number of PV components which are not easy to implement in systematic calculations.
Moreover, the limit of large PV masses has to be performed numerically. This may also be rather delicate to achieve in large systematic numerical calculations.

{\it iii) The dimensional regularization method}. This method is largely used in perturbation theory in the covariant Feynman approach. In LFD however, it has never been used since one would have to reformulate LFD in arbitrary D dimensions. This is also not in the spirit of LFD which deals only with physical degrees of freedom.

{\it iiv) The Bogoliubov-Parasiuk-Hepp-Zimmermann (BPHZ) method.} In this method each Feynman graph is treated separately. The
contribution is made finite by subtracting as many terms as necessary from the Taylor expansion at zero external momenta of the 
integrand. All Feynman integrals being convergent, no intermediate regularization is required, thereby showing that all  $n$-point
functions are effectively regularization independent. They only depend on the renormalisation conditions. However there are some complications  
for zero masses and despite its theoretical importance, the BPHZ scheme is not easy to deal with in practical calculations. The link between 
the TLRS and BPHZ schemes resides in the use of specific test functions equal to their Taylor remainder of any order.\\

We shall exhibit in this study the relationship between TLRS and the PV-type regularization procedure. We shall illustrate 
this relationship in the simple, but instructive,  calculation of the various components of the self-energy of a fermion in second 
order perturbation theory in the Yukawa model, paying particular attention to the restoration of rotational invariance.

The plan of the article is the following. We recall in Sec.~\ref{TLRS} the main features of the TLRS renormalization scheme. We then calculate in Sec.~\ref{CLFD} the self-energy in second order perturbation theory in the Yukawa model. We draw our conclusions in Sec.~\ref{conc}.

%%%%%%%%%%%%%%%%%%%%%%%%%%%%%%%%%%%%%%%%%%%%%%%%
\section{The Taylor-Lagrange renormalization scheme and Pauli-Villars subtraction} \label{TLRS}
 It is a common lore \cite{collins} that any  field $\phi(x)$ (taken here as a scalar field for simplicity)
 should be considered as an operator-valued distribution. This means that it should be defined by its application on test functions, 
 denoted by $\rho$, with well identified mathematical properties. In flat space, the physical field ${\varphi}(x)$ is thus given by \cite{grange}
\begin{equation} \label{conv}
\varphi(x) \equiv \int d^Dy \phi(y) \rho(x-y)\ ,
\end{equation}
in $D$ dimensions.
If we denote by $f$ the Fourier transform of the test function, we can further write ${\varphi}(x)$ in terms of creation and destruction operators, leading to
\begin{equation}
\!\varphi (x)\!=\!\!\int\!\frac{d^{D-1}{\bf p}}{(2\pi)^{{\bf (D-1)}}}\frac{f(\varepsilon_p^2,{\bf p}^2)}{2\varepsilon_p}
\left[a^+_{\bf p} e^{i{\bf p.x}}+a_{\bf p}e^{-i{\bf p.x}}\right],\ 
\end{equation}
with  $\varepsilon^2_p = {\bf p}^2+m^2$.  

From this decomposition, it is apparent that test functions should be attached to each fermion and boson field. Each 
propagator being the contraction of two fields should be proportional to $f^2$. In order to have a dimensionless argument for  $f$, 
we shall introduce an arbitrary scale $\Lambda$ to "measure" all momenta. $\Lambda$ can be any of the masses of the constituents. 
To deal with massless theories, we shall take some arbitrary value. The final expression of any amplitude 
should be independent of $\Lambda$. In CLFD, the test function is thus a function of $\frac{{\bf p}^2}{\Lambda^2}$ only. 

As recalled in \cite{grange}, the test function $f$ should have two important properties:

{\it i)} The physical field $\varphi(x)$ should be independent of the choice of the test function. This later should therefore be chosen 
among the partitions of unity (PU). It is a function of finite support which is $1$ everywhere except at the boundaries. This 
choice is also necessary in order to satisfy Poincar\'e invariance since, if $f$ is a PU, any power of  $f$, $f^n$, is also a PU. In the limit where the test function goes to $1$ over the whole space, we then have $f^n \to f$ and Poincar\'e invariance is recovered.

{\it ii)} In order to be able to treat in a generic way singular distributions of any type, the test function is chosen as a super regular test function (SRTF). It is a function of finite
extension - or finite support -  vanishing with all its derivatives at its boundaries, either in the ultraviolet (UV) or infrared (IR) domain.

Any physical amplitude is thus written in a schematic way like
\begin{equation} \label{ampli1}
{\cal A}=\int_0^\infty dX\ T(X)\  f(X)\ ,
\end{equation}
for a one-dimensional distribution. In this form, the amplitude ${\cal A}$ does not differ from the calculation using a cut-off procedure. In the UV domain for example,
 the cut-off, denoted by $H$, would correspond to the support of $f$, with $f(X \ge H)=0$. In order to go beyond the use of a naive 
cut-off, we should investigate the scaling properties inherent to the limit $X \to \infty$ since in this limit $\eta^2 X$ goes also
 to $\infty$, where $\eta^2$ is an arbitrary dimensionless scale. To do that, we shall consider as an example a distribution $T(X)$ leading
 to an $X$ integral diverging like $\log(H)$ in the absence of $f(X)$ and use the Lagrange formula written in the 
following form, in the UV domain~:
\begin{equation} \label{la3a}
f(aX)=-X\int_a^\infty\frac{dt}{t}  \partial_X \left[f(Xt)\right] \ ,
\end{equation}
for an arbitrary intrinsic scale $a$ which can be chosen positive if $T$ has no other singularity at finite $X$ .
This formula is an identity for any function $f$ which is a SRTF. In order to introduce the arbitrary scale $\eta^2$, we shall consider a
running boundary condition, i.e. a boundary condition which depends on the given variable $X$ using
\begin{equation} \label{running}
H(X)\equiv \eta^2 X g(X)\ ,
\end{equation}
up to an additive arbitrary finite constant irrelevant in the UV domain. Note that the support of the test function is the same in
the right- and left-hand sides of Eq.~(\ref{la3a}). This implies that $Xt\le H(X)$ for any argument $Xt$ of the test function (see
\cite{grange} for more details). 

The variable $X$ on which the running condition (\ref{running}) is applied should not be linked to any intrinsic scale. This is necessary in order to make sure that the limit $X \to \infty$ is properly done, i.e. that $X$ should be larger than any other physical scale present in the amplitude. Once this is done, the test function depends generally on $aX$, where $a$ is a priori a function of the kinematical variables of the system under consideration, and one should consider the Lagrange formula in the form (\ref{la3a}).

In order to extend the test function to $1$ over the whole space, we shall consider a set of function $g(X)$, denoted by $g_\alpha(X)$, where 
by construction $\alpha$ is a real positive number less than $1$. A typical
example of  $g_\alpha(X)$ is given in \cite{grange}, where it is shown that in the limit $\alpha \to 1^-$, with $\eta^2>1$, the running support of the PU test function
stretches then over the whole integration domain, and $f \to 1$. In this limit $g_\alpha(X) \to 1$.   

After integration by parts, the amplitude ${\cal A}$ writes 

\begin{equation} \label{afind}
{\cal A}=\int_0^\infty dX \ \partial_X \left[ X T(X)\right] \int_a^{\eta^2 g_\alpha(X)} \frac{dt}{t} f(Xt)\ .
\end{equation}
In the limit $\alpha \to 1^-$, the requirements are such that $g_\alpha(X) \to 1$ and $f \to 1$. One can thus define the extension in the UV domain, denoted by $\widetilde T^>$, of the singular distribution $T$ by
\begin{equation} \label{Ato}
{\cal A}\equiv \int_0^\infty dX \ \widetilde T^>(X)\ ,
\end{equation}
with\footnote{ Here and in Eq.(\ref{TIR2}) below the extensions for
$\widetilde{T}(X)$ are only valid when taken under the $X$-integral symbol, see \cite{grange} for discussions} 
\begin{equation} 
\widetilde T^>(X)\equiv \partial_X \left[ X T(X)\right] \mbox{Log}\left[\frac{\eta^2}{a}\right] \ ,
\end{equation}
where the derivative should be understood in the sense of distributions \cite{grange}.
It depends logarithmically on the arbitrary scale $\eta^2$, 
with $\eta^2>1$. The amplitude (\ref{Ato}) is now completely finite. 
Note that we do not need the explicit form of the test function in the derivation of the extended distribution $\widetilde T^>(X)$. 
We only rely on its mathematical properties  and on the running construction of the boundary conditions. 

The running boundaries of the test functions are essential for the preservation of symmetries which would
otherwise be destroyed with the usual cut-off test functions. Qualitatively, the reason for this property
is the following: the cut-off functions are equal to $1$ up to a point $x_c$ and fall then down to zero over a finite interval with some shape which does not change when $x_c$ is sent to infinity. On the contrary, with running boundaries, the width of the region where the test function falls from $1$ to zero increases proportionally to $x_c$ when $x_c$ goes to infinity, implying an infinitesimal drop-off of the test functions in the asymptotic limit. We call this behavior an ultrasoft cut-off.

The extension of singular distributions in the IR domain can be done similarly \cite{grange}. For an homogeneous distribution in one dimension, with $T[X/t]=t^{k+1} T(X)$, the extension of the distribution $T$ in the IR domain writes \footnote{At $d$ dimension, $k$ is defined by $T[X/t]=t^{k+d} T(X)$. This corrects a misprint in \cite{grange}.}

\begin{equation} \label{TIR2}
\widetilde T^<(X)=(-1)^{k}\partial_{X}^{k+1} \left[ \frac{X^{k+1}}{k!} T(X) \mbox{Log} (\tilde \eta X)\right] \ .
\end{equation}
The extension $\widetilde T^<(X)$ differs from the original distribution 
$T(X)$ only at the singularity at $X=0$.

The amplitude (\ref{afind}) can also be transformed alternatively in order to exhibit a PV-type subtraction. Using the Lagrange formula in the form
\begin{equation} \label{lat}
f\left[a X \right]=-\int_a^\infty dt \ \partial_t f\left[X t\right] \ ,
\end{equation}
we can rewrite the physical amplitude $\cal A$ in the following form, after the change of variable $Z=Xt$ and  in the limit
$\alpha \to 1^-$  
\begin{equation}
{\cal A}=-\int_0^\infty dZ  \int_a^{\eta^2} dt \  \partial_t \ \left[ \frac{1}{t} \ T\left( \frac{Z}{t} \right) \right] \ .
\end{equation}

With a typical  distribution $T(X)=\frac{1}{X+a}$ with an intrinsic scale  $a$, one thus gets immediately
\begin{equation}
\widetilde T^>(X) = \frac{1}{X+a} - \frac{1}{X+\eta^2}\ .
\end{equation}
We recover here a PV-type subtraction, with a scale $\eta^2$. This scale is completely arbitrary, with $\eta^2>1$ and 
not compulsory infinitely large as required for the PV masses. This extension of $T(X)$ leads to a well defined amplitude (\ref{Ato}).

%%%%%%%%%%%%%%%%%%%%%%%%%%%%%%%%%%%%%%%%%%%%%%%%
\section{Application to the calculation of the self-energy in light-front dynamics} \label{CLFD}
CLFD was first proposed in Ref.~\cite{karmanov} and detailed in the case of few-body systems in Ref.~\cite{cdkm}. Any physical system of momentum $p$ is described in LFD by a state vector projected onto the plan $t^+ = t + \frac{z}{c}$. In CLFD, the state vector is defined on the general plane determined by the equation $\sigma = \omega \cd x$, with $\omega ^2=0$.  The covariance of our approach
is due to the invariance of the light-front plane
equation. This implies that $\omega$ is not the
same in any reference frame, but varies according to Lorentz
transformations, like the coordinate $x$. It is not the case in
the standard formulation of LFD where $\omega$ is fixed to
$\omega=(1,0,0,-1)$ in any reference frame. The evolution of the system is thus defined in terms of the light-front time $\sigma$. 

We shall consider in the following the simple case of the self-energy of a fermion in the Yukawa model, in second order perturbation theory. From a practical point of view, any amplitude is calculated using the equivalence of Feynman rules, as detailed in Ref.~\cite{cdkm}. 

\begin{figure}[btph]
\begin{center}
\includegraphics[width = 18pc]{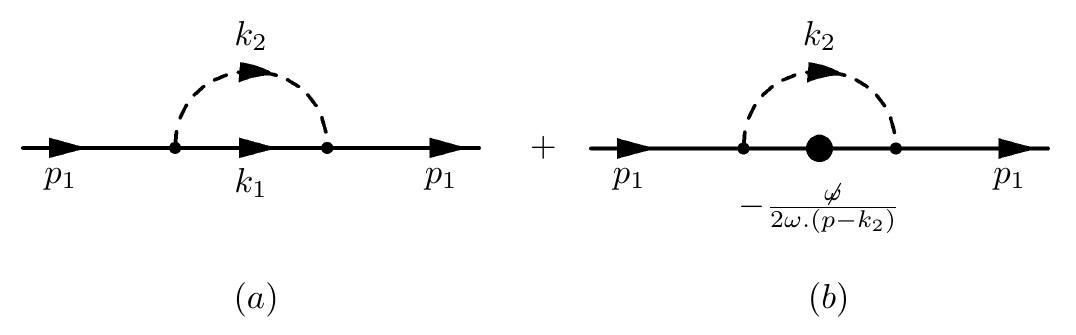}
\caption{\label{fig0a} Two contributions to the LFD fermion
self-energy $-\Sigma(p)$:  the  two-body intermediate state (a)
and the contact term (b). The solid and dashed lines
represent the fermion and the boson, respectively. }
\end{center}
\end{figure}

In second order of perturbation theory, the self-energy $\Sigma(p)$  (up to a conventional minus sign) is determined by the sum of the two
diagrams shown in Fig.~\ref{fig0a},
\begin{equation} \label{selfenLFD} 
\Sigma(p)=\Sigma_{2b}(p)+\Sigma_{fc}(p)\ .
\end{equation}
They correspond to the two-body contribution and the fermion
contact term, respectively.  These diagrams correspond to time ordered diagrams in the light-front time $\sigma$. Analytical expressions for the
corresponding amplitudes  read
\begin{multline}
\Sigma_{2b}(p)=-\frac{g^2}{(2\pi)^3}\int \delta^{(4)}(p+\omega\tau-k_1-k_2) \frac{d\tau}{\tau} \\
(\sla{k_1}+m)\theta(\omega \cd k_1)\delta(k_1^2-m^2) f^2\left[\frac{{\bf k_1}^2}{\Lambda^2} \right]  d^4k_1  \\
\theta(\omega\cd k_2)\delta(k_2^2-\mu^2) f^2\left[\frac{{\bf k_2}^2}{\Lambda^2} \right]  d^4k_2 \ ,
\label{2b}
\end{multline}
\begin{equation} \label{2c}
\Sigma_{fc}(p)=\frac{g^2}{(2\pi)^3} \sla{\omega} \int
\frac{\theta(\omega\cd
k_2)\delta(k_2^2-\mu^2)}{2\omega\cd(p-k_2)}  \\f^2\left[\frac{{\bf k_2}^2}{\Lambda^2} \right]d^4k_2\ ,
\end{equation}
where $g$ is the coupling constant of the fermion-boson
interaction, and $m$ and $\mu$ are the fermion and boson masses,
respectively.  Since the test functions are PU, we shall identify in the following $f^2$ with $f$. All the particles are on their mass shell in LFD, but off energy shell, so that the momentum $\tau$ represents the off-shell  energy of the intermediate fermion-boson state \cite{cdkm}. The momentum $p$ corresponds to $p=p_1-\omega \tau_1$ where $\tau_1$ is proportional to the off-shell energy of the initial, or final, fermion, and $p_1$ is the on-shell four-momentum of the fermion.

Note that in LFD, the self-energy may depend a priori on the position of the light-front characterized by $\omega$. In an exact calculation,
or in perturbation theory, we should check explicitly that this dependence disappears in order to recover the well known results in the 4D Feynman approach. 

The self-energy can therefore be decomposed in the most general spin structures according to
\begin{subequations}
\label{sigdec}
\begin{eqnarray}
\Sigma_{2b}(p)\!\!\!&=&\!\!\!g^2\left[{\cal A}(M^2)+{\cal B}(M^2)\frac{{\sla
p}}{M}+{\cal C}(M^2)\frac{M {\sla \omega}}{\omega\cd p}\right]
\label{Sigdecomp2b} \ ,\nonumber \\
\\
\Sigma_{fc}(p)\!\!\!&=&\!\!\!g^2C_{fc}\frac{M{\sla \omega}}{\omega\cd p}\ ,
\label{Sigdecompfc}
\end{eqnarray}
\end{subequations}
where the coefficients ${\cal A}$, ${\cal B}$, and ${\cal C}$ are
scalar functions which depend on $p^2$ only. They are independent
of $\omega$. We denote $p^2$ by $M^2$. The scale $M$ in (\ref{sigdec}) is just introduced for convenience in order to have the same dimension for all the coefficients ${\cal A},{\cal B},{\cal C}$ and $C_{fc}$. 

These coefficients can easily be calculated according to 
\begin{subequations}\label{eq4sen} 
\begin{eqnarray}
g^2{\cal A}(M^2)&=&\frac{1}{4}\mbox{Tr}[\Sigma(p)]\ , \\
g^2{\cal B}(M^2)&=&\frac{M}{4\omega\cd p}
\mbox{Tr}[\Sigma(p)\sla{\omega }]\ , \\
g^2{\cal C}(M^2)&=&\frac{1}{4M}\mbox{Tr}\left[\Sigma(p)\left(\sla{p}
-\frac{M^2\sla{\omega}}{\omega\cd p}\right)\right]\ .
\end{eqnarray}
\end{subequations}
The coefficient $C_{fc}$ is a constant. It can be extracted directly from Eq.~(\ref{2c}). 

In order to transform (\ref{2b}) using TLRS, we shall use the Lagrange formula (\ref{la3a}) in a slightly different form
\begin{equation} \label{latra}
f\left[a^2X^2 \right]=-\int_a^\infty dt \ \partial_t f\left[X^2 t^2\right] \ ,
\end{equation}
where the scale $a$ should be identified later on. With the change of variable  $\bar k_1 = k_1 s t$, $\bar k_2 = k_2 s t$, and $ \bar \tau = \tau s t$, we have
\begin{widetext}
\begin{multline} \label{s2b}
\Sigma_{2b}(p)=-\frac{g^2}{(2\pi)^3} \int d^4\bar k_1 \int d^4 \bar k_2 \int \frac{d \bar \tau}{ \bar \tau} \int_a^\infty dt \ \partial_t \int_a^\infty ds  \ \partial_s 
\frac{1}{st} \ 
\delta^{(4)}(p s t +\omega \bar \tau-\bar k_1-\bar k_2) \\
\theta(\omega\cd \bar k_2)\delta(\bar k_2^2-\mu^2 s t) 
(\sla \bar k_1 +m s t)\ \theta(\omega\cd \bar k_1)\delta(\bar k_1^2-m^2 s t )\  f\left[\frac{{\bf \bar k_2}^2}{s^2 a^2\Lambda^2} \right]   f\left[\frac{{\bf \bar k_1}^2}{t^2 a^2\Lambda^2} \right] \ .
\end{multline}
\end{widetext}
The four-momentum conservation law, and the on-mass shell conditions in Eq.~(\ref{s2b}) are equivalent to the original ones in (\ref{2b}) after the transformation 
\begin{equation} \label{mst}
 m,\mu,p \to mst, \mu st,pst \ .
 \end{equation}

Using the kinematical variables defined by
\begin{equation} \label{lf}
x=\frac{\omega \cd \bar k_2}{\omega \cd \ p s t} \ \ , \ \ R=\bar k_2-x p s t \ \ \mbox{with}
\ \ R=(R^0, {\bf R}_\perp, R^\parallel) \ ,
\end{equation}
we have, in the reference frame where ${\bf p} = 0$, ${\bf R}_\perp={\bf k}_\perp$. Since $\omega \cd R=0$, we also have $R^0=R^\parallel$. The momentum ${\bf k}_\perp$ is the perpendicular component of the four-momentum $\bar k_2$ with respect to the position $\omega$ of the light-front. With the transformations (\ref{mst}), we have thus, in the limit of large momenta (UV regime) and using the kinematics detailed in Appendix B.1 of \cite{grange}
\begin{eqnarray} \label{k2inf}
{\bf \bar k}_2^2 &\approx&  \frac{{\bf k}_\perp^4}{4x^2M^2 s^2 t^2}\ , \\
\label{k1inf}
{\bf \bar k}_1^2 &\approx &\frac{{\bf k}_\perp^4}{4(1-x)^2M^2 s^2 t^2}\ .
\end{eqnarray}
To simplify the notation, we shall define the dimensional scale
\begin{equation}
\alpha(x)=\frac{m^2x+\mu^2(1-x)-M^2x(1-x)}{m^2} \ .
\end{equation}

With the change (\ref{mst}), the coefficients ${\cal A}, {\cal B}$  and ${\cal C}$ thus write \cite{kms_04}
\begin{widetext}
\begin{subequations}
\label{ABC2}
\begin{eqnarray}
\label{Ap2} {\cal
A}(M^2)&=&-\frac{m g^2}{16\pi^2}\int_0^{\infty}d{\bf k}_{\perp}^2\int_0^1
dx\int_a^\infty \! \! \!dt \  \partial_t \!\int_a^\infty \!\! \! \!ds \ \partial_s \,\frac{1}{{\bf k}_{\perp}^2+m^2 s^2t^2\alpha(x)} f[] f[],\nonumber \\
\\
\label{Bp2} {\cal
B}(M^2)&=&-\frac{M g^2}{16\pi^2}\int_0^{\infty}d{\bf k}_{\perp}^2\int_0^1
dx\int_a^\infty \! \! \!dt \  \partial_t \!\int_a^\infty \!\! \! \!ds \ \partial_s\,\frac{(1-x)}{{\bf k}_{\perp}^2+m^2 s^2t^2\alpha(x)}f[] f[]\ ,\nonumber \\
\\
\label{Cp2} {\cal
C}(M^2)&=&-\frac{g^2}{32\pi^2M}\int_0^{\infty}d{\bf k}_{\perp}^2\int_0^1
dx\int_a^\infty \! \! \!dt \  \partial_t \!\int_a^\infty \!\! \! \!ds  \ \partial_s \ \frac{1}{s^2t^2}\,\frac{k_{\perp}^2+[m^2-M^2(1-x)^2]s^2t^2}{(1-x)[{\bf k}_{\perp}^2+m^2 s^2t^2\alpha(x)]} f[] f[] , \nonumber \\
\end{eqnarray}
\end{subequations}
\end{widetext}
where the notation $f[] f[]$ stands for 
\begin{multline}
f[] f[]=\ f\left[\frac{{\bf k}_\perp^4}{4x^2M^2 a^2 \Lambda^2 s^4 t^2} \right]   \\
f\left[\frac{{\bf k}_\perp^4}{4(1-x)^2M^2 a^2 \Lambda^2 s^2 t^4}\right]\ .
\end{multline}
The additional factors $st$ and $\frac{1}{st}$ in (\ref{Bp2}) and (\ref{Cp2}) respectively, as compared to (\ref{s2b}), originate from the momentum dependence of the self-energy in the decomposition (\ref{Sigdecomp2b}) with the replacem�ent $p \to p st$.

In order to calculate the coefficient  ${\cal A}$, we can separate the integration over $x$  in two domains, for $x<\frac{1}{2}$ and $x \ge \frac{1}{2}$, as already done in \cite{grange}. In the first domain, we can first eliminate the integral in $s$ by redefining $\tilde {\bf k}_\perp = \frac{ {\bf k}_\perp}{s}$, and using the Lagrange formula (\ref{latra}) on $s$ backward. It thus reads
\begin{multline} \label{afinal}
 {\cal A}(M^2)=\frac{m g^2}{16\pi^2} \int_0^{\infty}d{\bf \tilde k}_{\perp}^2  \int_0^\frac{1}{2}
dx \\
\int_a^\infty \! \! \!dt \  \partial_t \frac{1}{{\bf \tilde k}_{\perp}^2+m^2 t^2 \alpha(x)}\ f\left[\frac{{\bf \tilde k}_\perp^4}{4x^2M^2 a^2 \Lambda^2 t^2} \right]  \ ,
\end{multline}
where the second test function has been put to $1$ since its argument is always smaller than the one retained in (\ref{afinal}). With the change of variable $Z=\frac{{\bf \tilde k}_\perp^2}{2xM\Lambda t}$, we get
\begin{multline}
 {\cal A}(M^2)=\frac{m g^2}{16\pi^2}  \int_0^{\infty}dZ  \int_0^\frac{1}{2}
dx\int_a^\infty \! \! \!dt \  \partial_t  \\
\frac{2xM\Lambda}{m^2}\frac{1}{\frac{2xM\Lambda}{m^2} Z+ t \alpha(x)}\ f\left[\frac{Z^2}{a^2}\right]  \ .
\end{multline}
By eliminating the intrinsic scale $\alpha(x)$ with the identification $a=\alpha(x)$, and with $Z=a Y$, we get
\begin{multline}
 {\cal A}(M^2)=\frac{m g^2}{16\pi^2} \int_0^{\infty}dY  \int_0^\frac{1}{2}
dx \int_{\alpha(x)}^\infty \! \! \!dt \  \partial_t  \\
\frac{2xM\Lambda}{m^2}\frac{1}{\frac{2xM\Lambda}{m^2} Y+ t }\ f[Y^2]  \ .
\end{multline}
The integration over $t$ gives simply, with $f\to 1$ and the upper limit fixed by the running condition $Y^2t^2 \le H(Y^2)$, i.e. $t \le \eta$
\begin{multline}
 {\cal A}(M^2)=\frac{m g^2}{16\pi^2} \int_0^{\infty}dY  \int_0^\frac{1}{2} dx \\ \frac{2xM\Lambda}{m^2}\left[ \frac{1}{\frac{2xM\Lambda}{m^2}Y+\eta}  - \frac{1}{\frac{2xM\Lambda}{m^2}Y+\alpha(x)}\right]\ .
\end{multline}
The calculation in the interval $\frac{1}{2} \le x < 1$ and a first integration over $t$ instead of $s$ gives  the same integrand, so that we finally get
\begin{widetext}
\begin{eqnarray}
 {\cal A}(M^2)&=&\frac{m g^2}{16\pi^2} \int_0^{\infty}dY  \int_0^1 dx \frac{2xM\Lambda}{m^2}  \left[ \frac{1}{\frac{2xM\Lambda}{m^2}Y+\eta}  - \frac{1}{\frac{2xM\Lambda}{m^2}Y+\alpha(x)}\right]\nonumber \\
 &=&-\frac{m g^2}{16\pi^2} \log \eta+\frac{m g^2}{16\pi^2}   \int_0^1 dx \log \left[ \frac{m^2x+\mu^2(1-x)-M^2x(1-x)}{m^2}\right] \ .
\end{eqnarray}
\end{widetext}
This result is the same as the one already given in \cite{grange}.
The calculation of ${\cal B}$ proceeds in exactly the same manner, with just an extra factor $(1-x)$ in the integrand. 

We shall now concentrate on the calculation of the coefficient ${\cal C}$. Using (\ref{Cp2}), we can decompose ${\cal C}$ in two parts
\begin{widetext}
\begin{eqnarray}\label{C1}
 {\cal
C}(M^2)&=&-\frac{g^2}{32\pi^2 M}\int_0^{\infty}d{\bf k}_{\perp}^2 \int_0^{1}  dx\int_a^\infty \! \! \!dt \  \partial_t \!\int_a^\infty \!\! \! \!ds  \ \partial_s \,\frac{m^2 \partial_x \alpha(x)}{{\bf k}_{\perp}^2+m^2 s^2t^2\alpha(x)} f[] f[] \nonumber \\
&&-\frac{g^2}{32\pi^2M}\int_0^{\infty}d{\bf k}_{\perp}^2 \int_0^{1}  \frac{dx}{1-x}\int_a^\infty \! \! \!dt \  \partial_t \!\int_a^\infty \!\! \! \!ds  \ \partial_s \ \frac{1}{s^2t^2} f[] f[] \\
&\equiv&{\cal C}_1 + {\cal C}_2 \ . \nonumber
\end{eqnarray}
\end{widetext}
Let us first calculate  ${\cal C}_1$. With the change of variable $\alpha(x) s^2 t^2 =u$, we have
\begin{multline}
{\cal C}_1=-\frac{g^2}{32\pi^2M}\int_a^\infty \! \! \!dt \  \partial_t \!\int_a^\infty \!\! \! \!ds  \ \partial_s  \ \frac{1}{s^2t^2} \\
\int_0^{\infty}d{\bf k}_{\perp}^2\int_{\frac{\mu^2s^2t^2}{m^2}}^{s^2t^2} \frac{m^2 du}{ {\bf k}_{\perp}^2+m^2 u}  f[] f[] \ .
\end{multline}
The test functions provide the convergence of the integral in ${\bf k}_\perp^2$, as well as an upper limit, $\eta$, in the $t$ and $s$
integrations from the running condition on the test functions. The order of integrations can be changed at will and we
remark that it is legitimate at this stage to set the test functions to $1$ since, after integration over $u$ the integral in ${\bf k}_\perp^2$
is henceforth finite. We thus get
\begin{multline}
{\cal C}_1=-\frac{g^2}{32\pi^2M}\int_a^\eta \! \! \!dt \  \partial_t \!\int_a^\eta \!\! \! \!ds  \ \partial_s  \ \frac{1}{s^2t^2} \\
\int_0^{\infty}d{\bf k}_{\perp}^2\left[ \log\left[\frac{{\bf k}_{\perp}^2}{m^2}+s^2t^2\right]-\log\left[\frac{{\bf k}_{\perp}^2}{m^2}+\frac{\mu^2}{m^2}s^2t^2\right] \right]\ .
\end{multline}
After a change of variable $X=\frac{{\bf k}_{\perp}^2}{m^2}+s^2t^2$ in the first term, and $X=\frac{{\bf k}_{\perp}^2}{m�^2}+\frac{\mu^2}{m^2}s^2t^2$ in the second, we get
\begin{equation}
{\cal C}_1=-\frac{g^2 m^2}{32\pi^2M}\int_a^\eta \! \! \!dt \  \partial_t \!\int_a^\eta \!\! \! \!ds  \ \partial_s  \ \frac{1}{s^2t^2}  \int_{\frac{\mu^2s^2t^2}{m^2}}^{s^2t^2} dX \log X \ .
\end{equation}
which finally gives, with $X=Y s^2t^2$
\begin{equation}
{\cal C}_1=-\frac{g^2 m^2}{32\pi^2M}\int_a^\eta \! \! \!dt \  \partial_t \!\int_a^\eta \!\! \! \!ds  \ \partial_s   \int_{\frac{\mu^2}{m^2}}^{1}dY \log [Ys^2t^2] \ .
\end{equation}
The final integrations over $s$ and $t$ lead to a double PV-type subtraction
\begin{multline}
{\cal C}_1=-\frac{g^2 m^2}{32\pi^2M} \int_{\frac{\mu^2}{m^2}}^{1}dY 
\left[\log [Y] - \log \left[Y\frac{\eta^2}{a^2}\right] \right. \\
\left. - \log \left[Y\frac{\eta^2}{a^2}\right]+ \log \left[Y\frac{\eta^4}{a^4}\right]\right] \equiv 0\ .
\end{multline}
This result is very similar to the calculation of the coefficient ${\cal C}$ in \cite{kms_04} where it was shown that it is zero with one PV fermion and one PV boson subtraction. These subtractions are here provided by the integration over the variables $s$ and $t$. Note that we indeed need both subtractions to get the final result. 

This result relies on the property that the numerator of the integrand in ${\cal C}_1$ in (\ref{C1})  is just the derivative of $\alpha(x)$.
This is not the case for any other components. Because of this peculiarity, the calculation of ${\cal C}_1$ proposed in \cite{grange}
(coefficient ${\cal J}_2$ in Appendix B.3) is only correct for this case since otherwise it would lead to a null result
independently of the form of the singular distribution. In the calculation of \cite{grange}, the identification of the product of the two test functions 
with only one symmetric in the change $x \to 1-x$ is legitimate since  under this change only the boundaries of the $u$-integral are
interchanged thereby giving zero for ${\cal J}_2$.

We turn now to the calculation of the coefficient ${\cal C}_2$. Since the integrand has no intrinsic scale, it is natural to calculate ${\cal C}_2$ using the IR extension given in (\ref{TIR2}) for an homogeneous distribution. With the change of variable $\frac{{\bf k}_\perp^2}{s^2t^2}=\frac{1}{Y}$, we can integrate over $s$ and $t$ using the Lagrange formula (\ref{la3a}) back. We get
\begin{multline}
 {\cal C}_2(M^2)=-\frac{g^2}{32\pi^2M} \int_0^{\infty}\frac{dY}{Y^2} \int_0^{1}  \frac{dx}{1-x} \\
 f\left[ \frac{1}{4x^2Y^2M^2\Lambda^2}\right] f\left[ \frac{1}{4(1-x)^2Y^2M^2\Lambda^2}\right]\ .
\end{multline}
To keep the symmetry $x \to (1-x)$ in the integrand, we shall rewrite ${\cal C}_2$ as
\begin{multline}
 {\cal C}_2(M^2)=-\frac{g^2}{32\pi^2M} \int_0^{\infty}\frac{dY}{Y^2} \int_0^{\frac{1}{2}}  \frac{dx}{x(1-x)} \\
 f\left[ \frac{1}{4x^2Y^2M^2\Lambda^2}\right] f\left[ \frac{1}{4(1-x)^2Y^2M^2\Lambda^2}\right]\ .
\end{multline}
In the domain $x \le 1/2$, the first test function only matters, i.e. is different from $1$ within the relevant  boundary. We have therefore, with the change of variable $2YxM\Lambda = u$
\begin{equation}
 {\cal C}_2(M^2)=-\frac{g^2 \Lambda}{16\pi^2}\int_0^{\frac{1}{2}}  \frac{dx}{1-x} \int_0^{\infty}\frac{du}{u^2} f\left[ \frac{1}{u^2}\right]\ .
\end{equation}
The test function just provides the extension of the distribution $\frac{1}{u^2}$ at $u=0$ from (\ref{TIR2}), which is the pseudo-function of $1/u^2$ \cite{grange}. We finally have
\begin{equation}
 {\cal C}_2(M^2)=-\frac{g^2 \Lambda}{16\pi^2} \log 2 \int_0^{\infty}du \ \mbox{Pf}\left[\frac{1}{u^2} \right] \equiv 0\ .
\end{equation}
We find that the coefficient $C(M^2)$ is identically zero, as it should be. 

The calculation of the contribution from the contact interaction can be done very similarly to the calculation of $C_2$. We have, from (\ref{2c}) and (\ref{Sigdecompfc}) \cite{kms_04}
\begin{equation}
C_{fc}=\frac{g^2}{32\pi^2M} \int_0^{\infty}d {\bf k}_{\perp}^2 \int_0^{\infty}  \frac{dx}{x(1-x)} \\
f\left[\frac{{\bf k}_\perp^4}{4x^2M^2 s^4 t^2} \right] \ .
\end{equation}
The integration over $x$ can be decomposed in two parts, for $x\le \frac{1}{2}$ and for $x > \frac{1}{2}$. In the first one, with  ${\bf k}_{\perp}^2 =\frac{1}{Y}$, we get $C_{fc}=-{\cal C}_2(M^2)=0$. The second one is identically zero with a principal value prescription  to calculate the integral at x=1.
This insures that the self-energy $\Sigma(p)$ is indeed independent of the arbitrary position of the light-front.

%%%%%%%%%%%%%%%%%%%%%%%%%%%%%%%%%%%%%%%%%%%%%%%%
\section {Conclusions} \label{conc}
We have shown in this study that the recently proposed TLRS \cite{grange} leads naturally to a regularization of any amplitude very similar to a PV-type subtraction. However, contrary to the original PV regularization method, the TLRS does not necessitate to perform any infinite mass scale limit.

The application of our formalism to the calculation of the self-energy in second order perturbation theory in the Yukawa model, using LFD, is very instructive. It is very similar to the standard calculation using PV fields with a  negative norm \cite{kms_04}. The coefficients  
${\cal A}$ and ${\cal B}$ in the spin decomposition (\ref{Sigdecomp2b}) depends in TLRS on an arbitrarily dimensionless scale $\eta$, while it depends on $\frac{\Lambda_{PV}^2}{m^2}$ in the PV method, where $\Lambda_{PV}$ is the PV boson mass. This mass must be taken very large compared to any physical mass scale present in the amplitude.  In both methods, the regularization of the amplitude is achieved by using a single PV-type subtraction. On the other hand, the coefficient ${\cal C}$ is identically zero, as required by rotational invariance. It is regularized in both methods by using two PV-type subtractions, involving both the fermion and boson propagators.

This close connection between TLRS and PV-type regularization is of particular interest in further nonperturbative calculations in LFD \cite{kms_08}. In the TLRS scheme indeed, there is no need for additional non physical components in the state vector describing any physical system. Moreover, there is no large mass scale limit to perform numerically. This may render possible large scale calculations of nonperturbative relativistic bound state systems in LFD.

%%%%%%%%%%%%%%%%%%%%%%%%%%%%%%%%%

\end{document}